\documentclass[aip,reprint]{revtex4-2}
\usepackage{epsfig,amssymb,amsmath}

\usepackage{ulem}

\draft 

\begin{document}


\title{Resonant control of magnetization in a shunted $\varphi_0$ junction with LC circuit} 




\author{I. R. Rahmonov}
\affiliation{BLTP, JINR, Dubna, 141980, Moscow Region, Russia}
\affiliation{Dubna State University, Dubna, 141980, Moscow Region, Russia}
\affiliation{Moscow Institute of Physics and Technology, Dolgoprudny, 141700, Moscow Region, Russia}

\author{Yu. M. Shukrinov}
\affiliation{BLTP, JINR, Dubna, 141980, Moscow Region, Russia}
\affiliation{Dubna State University, Dubna, 141980, Moscow Region, Russia}
\affiliation{Moscow Institute of Physics and Technology, Dolgoprudny, 141700, Moscow Region, Russia}

\author{O. A. Kibardina}
\affiliation{Dubna State University, Dubna, 141980, Moscow Region, Russia}

\author{S. A. Abdelmoneim}
\affiliation{Menoufia University, Faculty of Science, Physics Department, Shibin al Kawm, 32511, Egypt}

\date{\today}

\begin{abstract}
The possibility of magnetization resonant control in a Josephson superconductor-ferromagnet-superconductor $\varphi_{0}$ junction shunted by an $LC$ circuit is demonstrated. As a result of the resonance of Josephson oscillations with oscillations in the circuit, a time-independent superconducting current arises in the junction. 
Due to the coupling of the Josephson phase and the magnetization of the ferromagnetic layer, the resulting superconducting current leads to a deviation of the easy axis from its initial position and to a precession of the magnetization around the tilted axis. We show that the tilt value increases with the increasing spin-orbit interaction and the Josephson to magnetic energy ratio. An analytical expression for the magnetization tilt is obtained, which agrees well with the results of numerical calculations. The emerging possibility of resonant control of magnetization in a shunted $\varphi_{0}$ junction can be used in the development of novel  technologies in the field of superconducting electronics and spintronics.
\end{abstract}
\keywords{Superconducting electronics, superconducting spintronics, $\varphi_0$-junction, spin-orbit interaction, shunted Josephson junction, resonance branch}

\maketitle

The presence of a phase shift in the current - phase relationship of a superconductor-ferromagnet-superconductor $\varphi_{0}$ junction associated with magnetization leads to the possibility of mutual influence of magnetization and superconducting current and is currently being intensively studied in connection with the prospects of applications in superconducting spintronics~\cite{golubov17,linder15,buzdin-prl08,konschelle-PRL,zhu17,houzet08,petkovic09,cai10,konschele15,chud16,Shukrinov_apl2017,borovets2018,Shukrinov-PRB,Shukrinov_ufn}. In particular, the resonance properties and the manifestation of the Kapitsa pendulum features in the $\varphi_{0}$ junction were considered in Refs~\cite{Shukrinov-PRB,Shukrinov_prb2021,Shukrinov_beilshtein,Shukrinov_epl2018}, the phenomenon of magnetization reversal by a current pulse and a magnetic field pulse was discussed in the works~\cite{Shukrinov_apl2017,rahmonov_pepanl2023}, 
the possibility of controlling a nanomagnet using superconducting current was demonstrated in the work~\cite{chud17}. C. Guarcello and F.S. Bergeret suggested a variant of creating cryogenic memory based on the $\varphi_{0}$ junction~\cite{Guarcello_pra}.

One of the effective ways to influence the properties of the Josephson structure is to shunt it with $LCR$ elements~\cite{hadley87,chernikov95,wiesenfeld96,cawthorne98,filatrella00,grib02,fistul07}. Shunting allows for effective control and manipulation of resonant features, which are used in superconducting electronics~\cite{Tolpigo}. The features of resonance phenomena and their manifestations on the current-voltage characteristics (CVC) of shunted Josephson junctions were reported in a number of experimental and theoretical works~\cite{Jensen,Larsen,Tachiki,Zhou,Shukrinov12,Licharev,Almaas11,Shukrinov15,Kornev,Jeanneret,Rfenacht,Welp,Ma,
Atanasov,Shukrinov17}.

In this paper, we propose a novel method for controlling magnetization based on shunting of the $\varphi_{0}$ junction with an $LC$ circuit. In addition to the existing ferromagnetic resonance, the so-called parallel resonance of Josephson oscillations with the oscillations of the resonant circuit arises~\cite{Licharev}.
In this case, an increase in the superconducting current is observed along the corresponding resonant branch on the CVC. We demonstrate that by changing the shunt parameters, it is possible to effectively influence the resonant properties of the Josephson junction, change the magnitude of magnetization and control its dynamics.

\begin{figure}[h!]
\centering
\includegraphics[width=8cm]{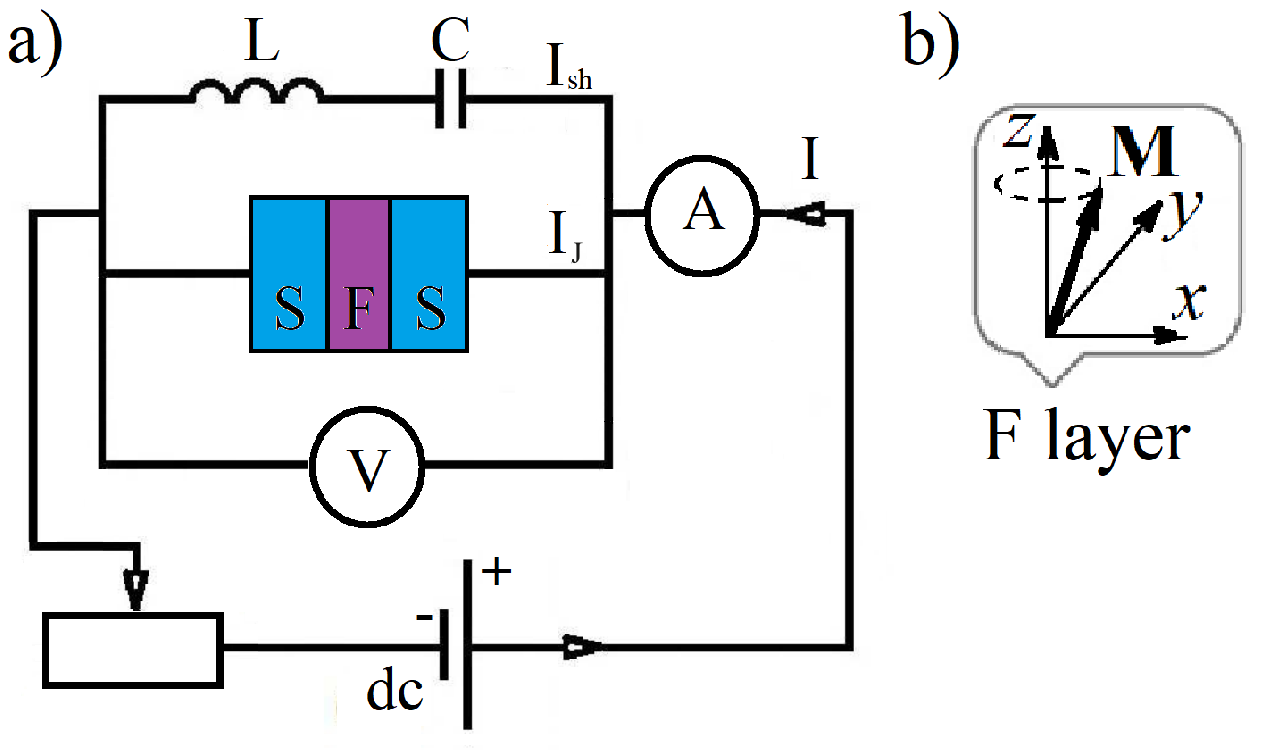}
\caption{(a) Scheme of the $\varphi_{0}$ junction shunted by LC elements, where S and F denote the superconducting and ferromagnetic layers, respectively; (b) Geometry of the ferromagnetic layer, where $\mathbf{M}$ is the magnetization vector. The easy axis of the ferromagnet is directed along the $z$ axis.}
\label{scheme}
\end{figure}

The scheme of the $\varphi_{0}$ junction shunted by the LC circuit and the geometry of the ferromagnetic layer are shown in Fig.\ref{scheme} (a) and (b), respectively. The easy axis of the ferromagnetic layer is directed along the $z$ axis.

The dynamics of magnetization $\mathbf{M}$ in the $\varphi_{0}$ Josephson junction (JJ) is described by the Landau-Lifschitz-Gilbert (LLG) equation
\begin{eqnarray}
\label{llg}
\frac{d\mathbf{M}}{dt}&=&-\gamma[\mathbf{M}\times\mathbf{H_{eff}}]+\frac{\alpha}{M_{0}}[\mathbf{M}\times\frac{d\mathbf{M}}{d t}],
\end{eqnarray}
with an effective magnetic field
\begin{eqnarray}
\label{heff2}
\mathbf{H_{eff}}=\frac{K}{M_{0}}\bigg[Gr\sin\bigg(\varphi-r\frac{M_{y}}{M_{0}}\bigg)\mathbf{e_{y}}+\frac{M_{z}}{M_{0}}\mathbf{e_{z}}\bigg]
\end{eqnarray}
Here $\gamma$ is the gyromagnetic ratio, $\alpha$ is the Gilbert damping and $M_{0}$ is the magnetization saturation value ($|\mathbf{M}|=M_{0}$), $\displaystyle G=\frac{E_{J}}{K\nu}$ is the ratio of the Josephson energy to the magnetic energy, and $\mathbf{e_{y}}$ and $\mathbf{e_{z}}$ are the unit vectors. Derivation of effective field expression is given in the Supplementary materials.

For investigations of the CVC and the dynamics of magnetization of the $\varphi_{0}$ junction, the LLG equations must be supplemented by the Josephson relation
\begin{eqnarray}
\label{jr}
V=\frac{\hbar}{2e}\frac{d\varphi}{d t},
\end{eqnarray}
expression for total current in the RCSJ-model
\begin{eqnarray}
\label{total_cur}
I&=&C_{J}\frac{d V}{d t}+\frac{\hbar}{2eR_{J}}
\bigg(\frac{d \varphi}{d t}-\frac{d \varphi_{0}}{d t}\bigg)\nonumber\\
&+&I_{c}\sin(\varphi-\varphi_{0})+C\frac{d u_{c}}{dt},
\end{eqnarray}
and by the equation for voltage determined by the equality of the sum of voltages of the shunted inductance $\displaystyle u_{L}=LC\frac{d^2 u_{c}}{dt^2}$, shunted capacitance $u_{c}$, and the voltage $V$ across the JJ, i.e.

\begin{eqnarray}
\label{uc}
LC\frac{d^2 u_{c}}{dt^2}+u_{c}=V,
\end{eqnarray}
where $C$ and $L$ are the shunt capacitance and inductance, respectively. The notation $C_J$, $R_J$, and $I_c$ is related to the capacitance, resistance, and critical current of the JJ.

Thus, equations (\ref{llg})-(\ref{uc}) form a coupled system of equations for describing the dynamics of the $\varphi_{0}$ junction shunted by an LC circuit. We note that in order to numerical solution of system of equations we have normalized all variables and we have introduced the following notations: $m_{i}$ is the magnetization component, normalized to magnetization saturation $M_{0}$, $t$ is time, normalized to the inverse Josephson plasma frequency $\omega_{p}^{-1}$, $\omega_{F}$ is ferromagnetic resonance frequency, normalized to 
$\omega_{p}$, $C$ is shunting capacitance, normalized to the Josephson capacitance $C_{J}$, $L$ is shunting inductance, normalized to $(\omega_{p}^{2}C_{J}))^{-1}$. Bias current $I$, is normalized to the critical current $I_{c}$, $V$ and $u_{c}$ are normalized to the $V_{0}=\hbar\omega_{p}/2 e$ and $\beta$ is is dissipation parameter of JJ. A full system of equation in normalized variables are given in the Supplementary materials. The $\varphi_{0}$ junction together with its shunted capacitance $C$ and inductance $L$ form an oscillatory circuit, where the resonance of Josephson oscillations with the oscillations of the resonant circuit with its eigenfrequency 
\begin{equation}
\label{w_par}
\omega_{rc}=\sqrt{\frac{C+1}{LC}},
\end{equation}
can be realized. We should note that in the case of resonance the voltage amplitude increases. To calculate the CVC, the system of equations (\ref{llg})-(\ref{uc}) is solved numerically using the fourth-order Runge-Kutta method.

\begin{figure}[h!]
\centering
\includegraphics[width=6cm]{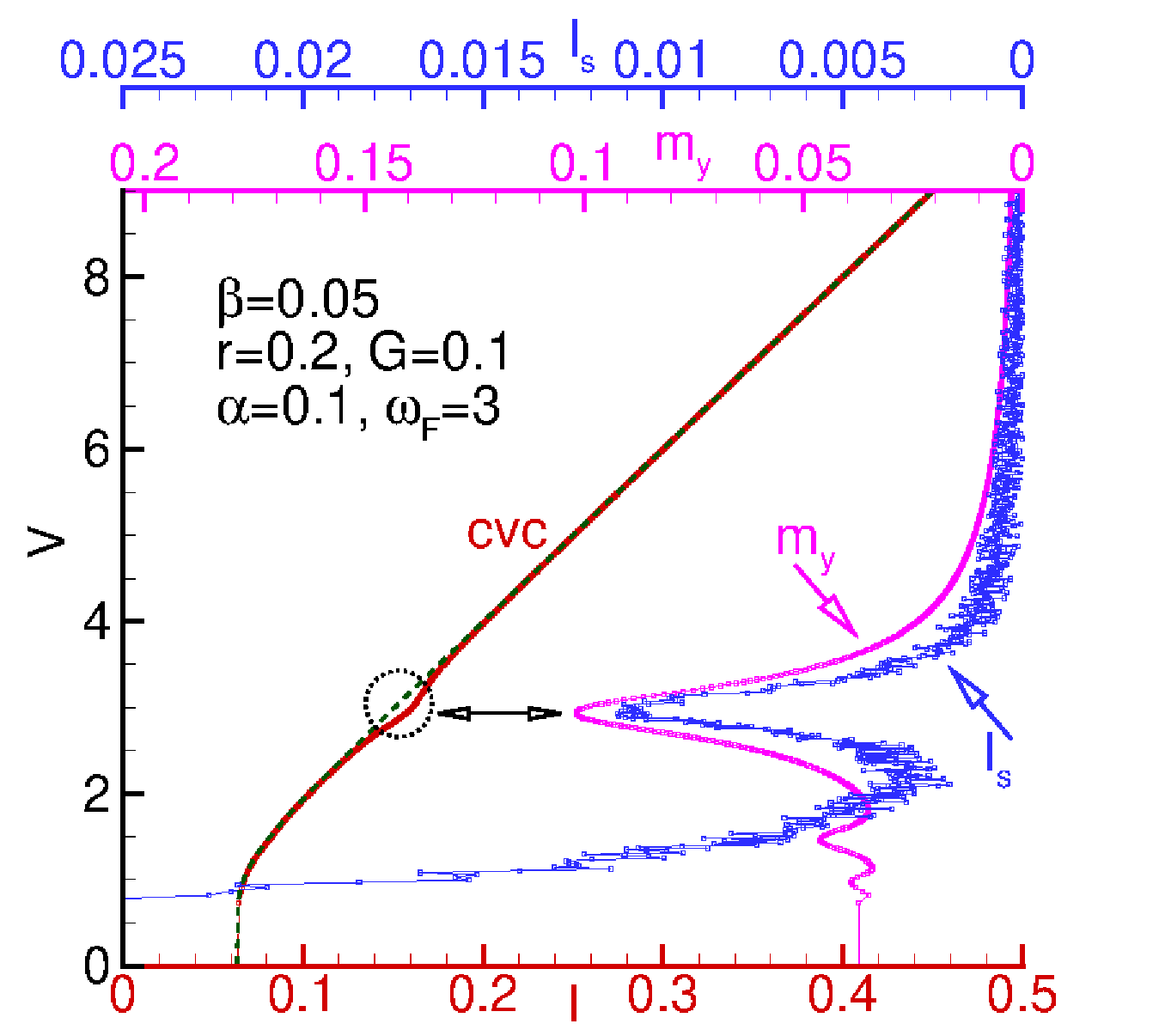}
\caption{The CVC of the $\varphi_{0}$ junction (red curve) together with the dependence of the average superconducting current (blue curve) and the amplitude $m_{y}$ (pink curve) on voltage, calculated for $\alpha = 0.1$, $G = 0.1$, $r = 0.2$, $\omega_{F}=3$ and $\beta = 0.05$. For comparison, the CVC of the SIS junction (dashed black curve) calculated for the dissipation parameter $\beta = 0.05$ is also shown.}
\label{phi0_jj}
\end{figure}

In order to highlight the effect of shunting on the system characteristics and make the obtained results more expressive, We first briefly consider the dynamics of the $\varphi_{0}$ junction. The calculated CVC together with the dependences of the superconducting current and the maximum value of the $m_y$ component of magnetization on voltage are shown in Fig. \ref{phi0_jj}. This figure also shows the CVC of the 
junction with insulating barrier (SIS, dashed line) for comparison. In the voltage region near $V=\omega_{F}=3$, the CVC of the $\varphi_{0}$ junction exhibits a dip structure in comparison with the CVC of the SIS junction, which is a manifestation of ferromagnetic resonance. In the region of ferromagnetic resonance $\omega_{F}=\omega_{J}$, under the influence of magnetization precession, a time-independent superconducting current arises, which leads to a decrease in the differential resistance, and manifestation of the above-mentioned dip structure~\cite{Shukrinov-PRB}. This fact is also demonstrated by the dependence of the maximum magnetization component $m_y$ on voltage, where its growth and the greatest value are observed at $\omega_{F}=\omega_{J}$.

\begin{figure}[h!]
\centering
\includegraphics[width=6cm]{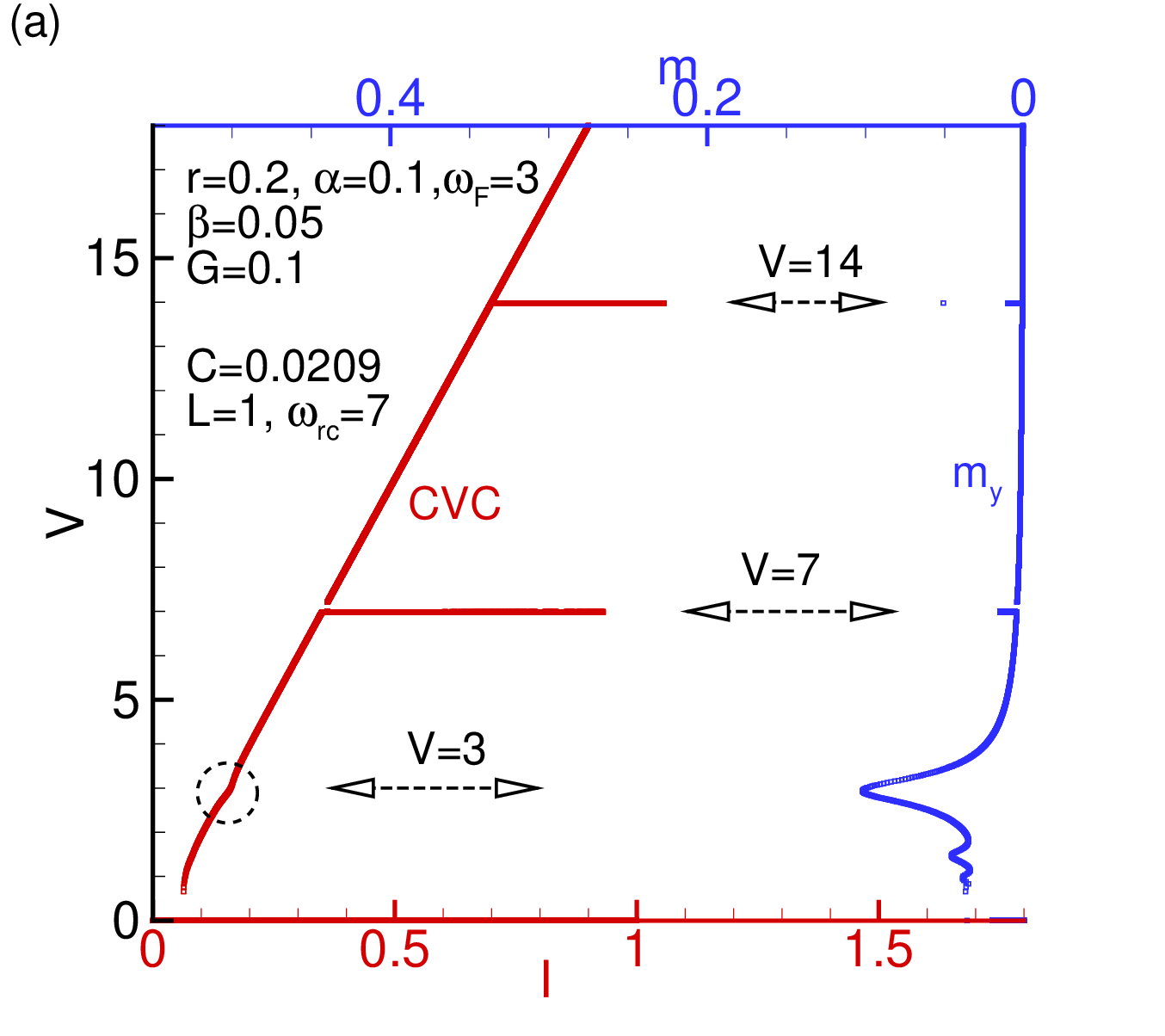}
\includegraphics[width=6cm]{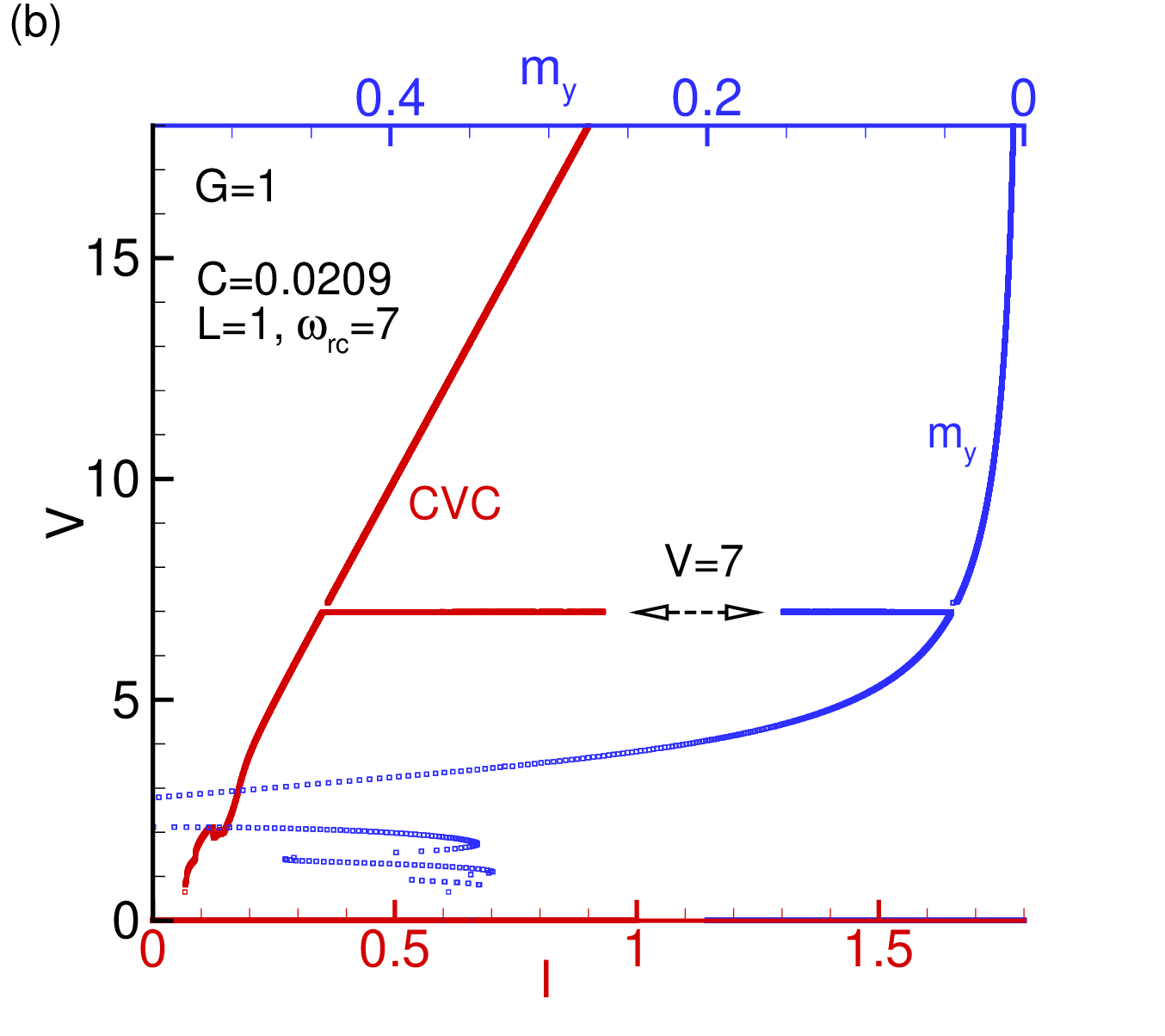}
\caption{
The CVC of the $\varphi_{0}$ junction shunted by an LC circuit (red curve) together with the dependence of the maximum value of $m_{y}$ (blue curve) on voltage calculated at $\beta=0.05$, $\alpha = 0.1$, $r = 0.2$, $\omega_{F}=3$, $C = 0.125$, $L = 1$ ($\omega_{rc}=3$); (a) $G = 0.1$; (b) $G = 1$. The double arrows indicate the correlation between the position of the resonant branch and the maximum value of $m_{y}$.}
\label{shunted_phi0}
\end{figure}

Now we consider the shunted $\varphi_{0}$ junction. The calculated CVC together with the dependences of $m_{y}^{max}$ on voltage are shown in Fig.\ref{shunted_phi0}(a). The selected values of capacitance ($C=0.0209$)and inductance ($L=1$) correspond to the resonant frequency $\omega_{rc}=7$. The calculations were performed with the same values of the model parameters as in the unshunted case presented above. The three-loop CVC demonstrates a resonant branch at voltage $V=\omega_{rc}=7$ and a branch corresponding to the second harmonic at $V=2\omega_{rc}=14$ (a detailed description of the calculation of the multi-loop CVC is given in the Supplementary materials).
The presented dependence of $m_{y}^{max}$ on voltage along the CVC demonstrates maxima in the case of the realization of ferromagnetic resonance ($V=\omega_{F}$) and its subharmonic $V=\omega_{F}/2$. It also shows an increase in $m_{y}^{max}$, correlating with the positions of the rc-branch and its harmonics on the CVC.

\begin{figure}[h!]
\centering
\includegraphics[width=6cm]{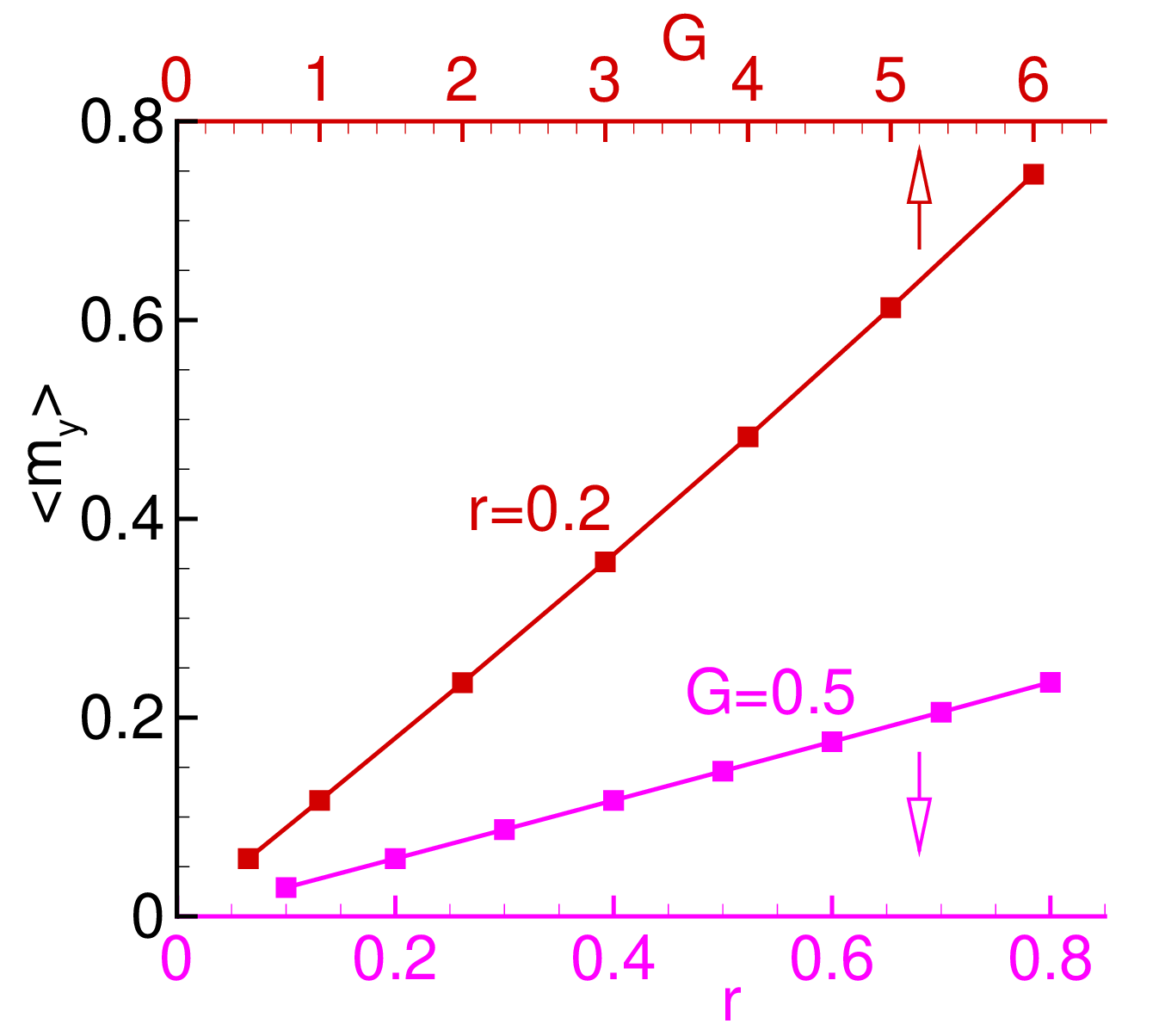}
\caption{
Dependence of the average value of $<m_{y}>$ at the point of the resonance maximum in the LC circuit on the the Josephson to magnetic energy ratio $G$ at $r=0.2$ and on the value of the spin-orbit coupling parameter $r$ at $G=0.5$. The calculations were performed for the model parameter values $\beta=0.05$, $\alpha = 0.1$, $\omega_{F}=3$, $C = 0.0209$, $L = 1$. The arrows indicate the curve's belonging to the corresponding axes.}
\label{Gdep}
\end{figure}

\begin{figure}[h!]
\centering
\includegraphics[width=7cm]{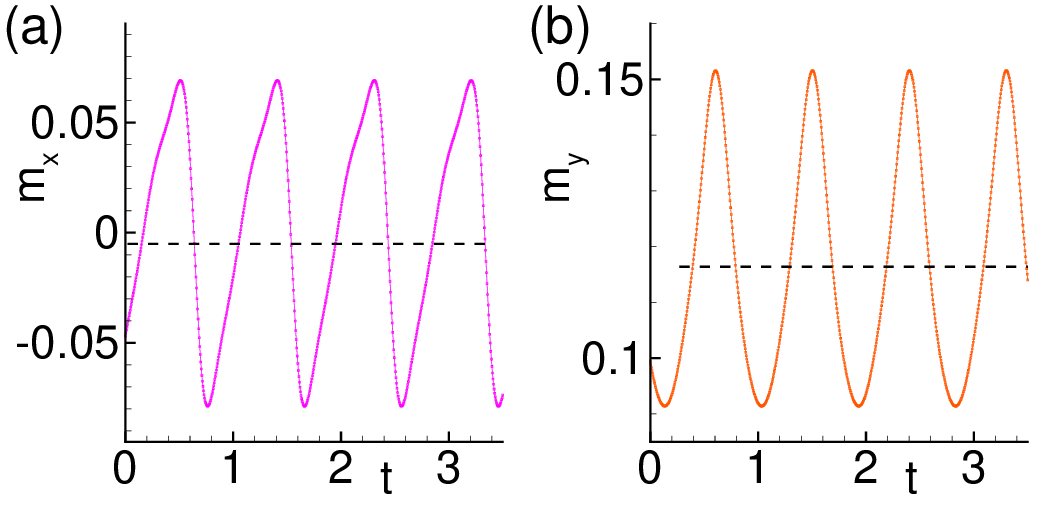}\\
\includegraphics[width=3.5cm]{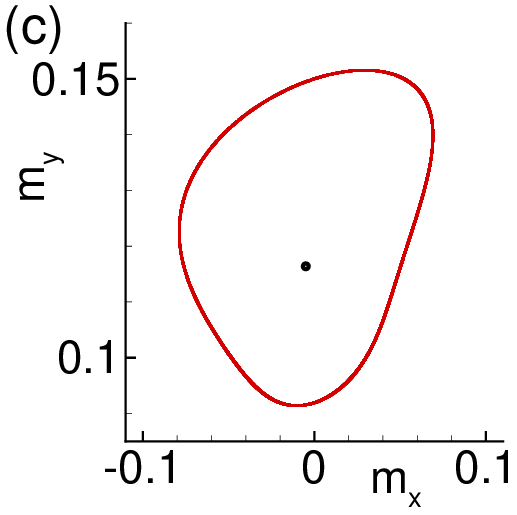}
\includegraphics[width=3.5cm]{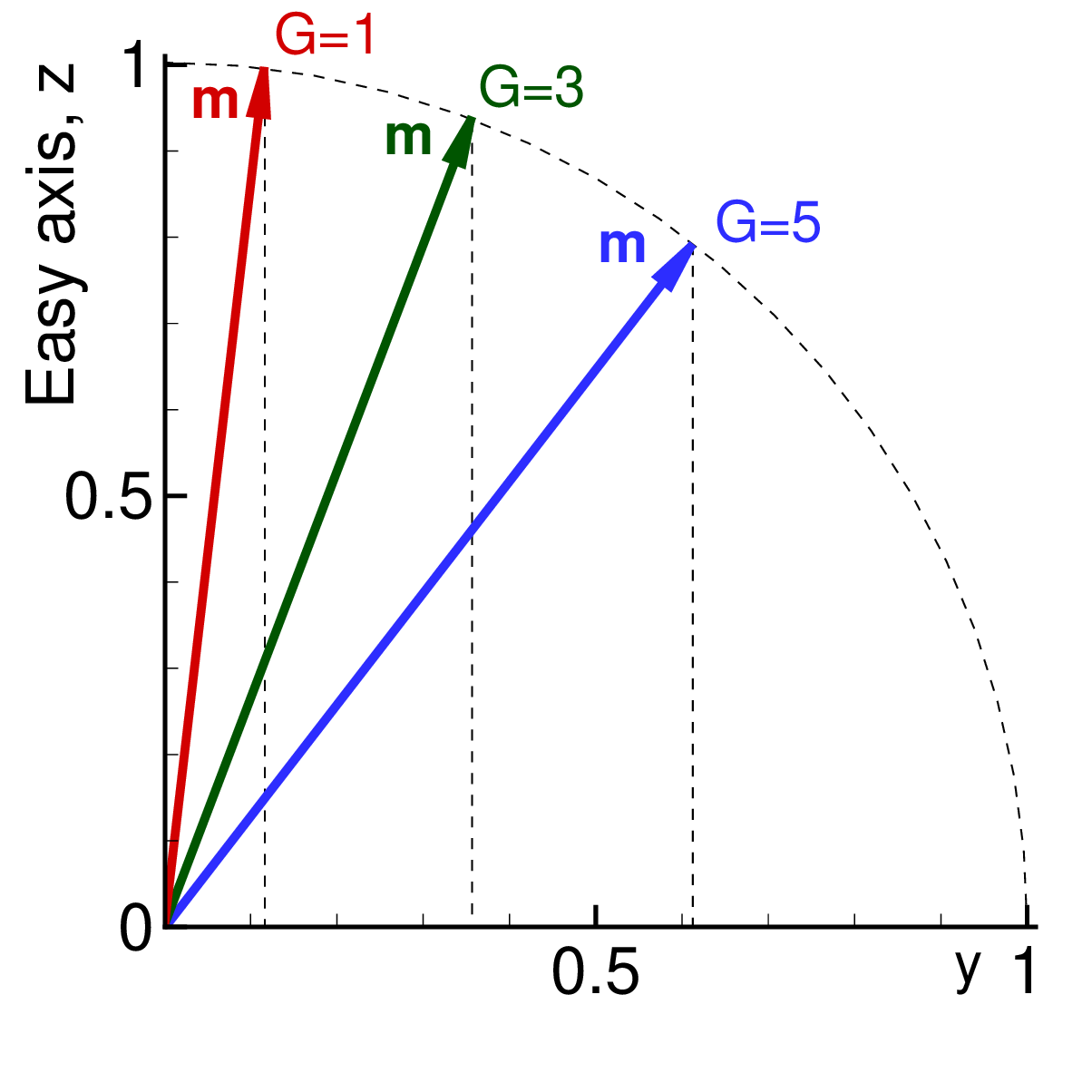}
\caption{Time dependence of magnetization components at the bias current value $I=0.93$: (a) $m_{x}$; (b) $m_{y}$; (c) the corresponding trajectory on the plane $m_{y}$-$m_{x}$; (d) Variation of the magnetization direction at different G.}
\label{time_dep}
\end{figure}

\begin{figure}[h!]
\centering
\includegraphics[width=6cm]{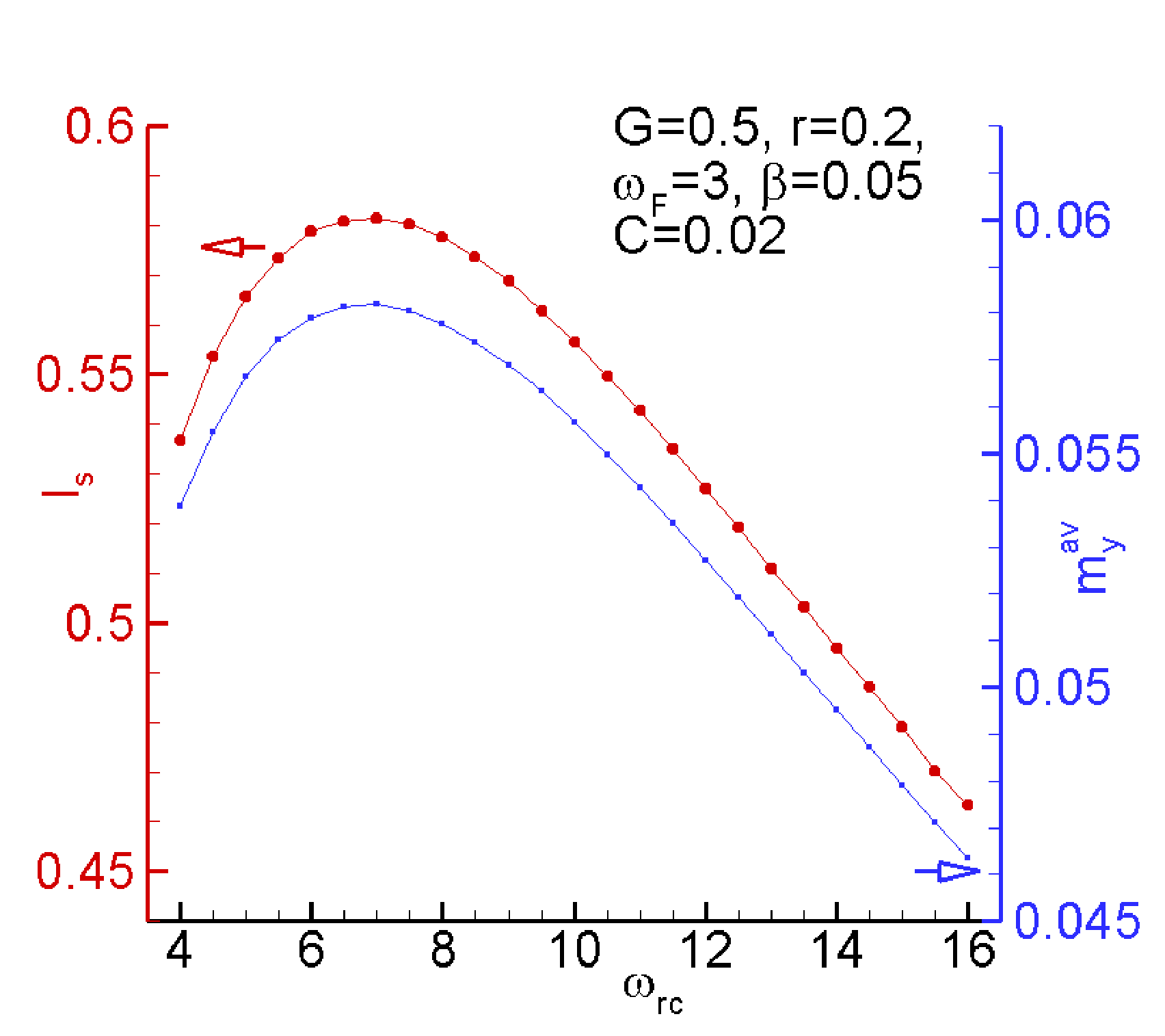}
\caption{Dependence of the average values of the magnetization component $m^{av}_{y}$ and the magnitude of the superconducting current on the frequency of the resonant circuit at $\beta=0.05$, $\alpha = 0.1$, $G = 5$, $r = 0.2$, $\omega_{F}=3$, $C = 0.02$ and different values of the shunt inductance $L$.}
\label{omega_dep}
\end{figure}

As usual, in the ferromagnetic resonance region, the magnetization precesses around the easy axis (in our case the z-axis). However, the presented results demonstrate an additional increase in $m_y^{max}$ in the region of the CVC corresponding to the parallel resonance region in the circuit, i.e. a deviation of the magnetization from the easy axis $z$ is observed. The results presented below (see Figs.\ref{time_dep} and \ref{omega_dep}) show a deviation towards to the $y$-axis. The calculations also show that with an increase in the ratio of the Josephson to magnetic energy $G$ and the spin-orbit interaction $r$, the magnitude of this deviation increases. This can be seen in the dependences of the average magnetization on the parameters $G$ and $r$, shown in Fig.\ref{Gdep}. Note that the nature of the increase of $m_{y}^{max}$ is linear with increasing $r$ and $G$ in the interval under consideration.

As noted above, the realized parallel resonance in the shunted $\varphi_{0}$ junction results in the appearance of a time-independent average superconducting current, which contributes to the effective magnetic field. This contribution leads to a deviation of magnetization from the easy axis toward the $y$ axis. This effect provides a unique opportunity to control magnetization by means of the resonance in the shunted $\varphi_{0}$ junction. We emphasize that the magnetization is controlled by changing the frequency of the LC circuit. We also note that the observed magnetization deviation can be registered using a DC-SQUID, which allows direct experimental investigations of the observed phenomenon.

In order to analytically confirm this effect, we consider a special case of the system of equations in the presence of only a time-independent superconducting current, which corresponds to the stationary Josephson effect regime, i.e., when $I<I_{c}$. In this case, Josephson oscillations are absent ($V(t)=0$), and therefore the magnetization components remain constant in time ($\mathbf{M}=const$) and the damping term is zero since $\frac{d\mathbf{M}}{dt}=0$. Taking into account these simplifications and the expressions for the effective field, the system of equations can be rewritten as:

\begin{eqnarray}
\label{syseq_simple}
m^{c}_{z}[m^{c}_{y}-rG\sin(\varphi-r m^{c}_{y})]=0\nonumber\\
m^{c}_{x}m^{c}_{z}=0\nonumber\\
m^{c}_{x}rG\sin(\varphi-r m^{c}_{y})=0\\
I-\sin(\varphi-r m^{c}_{y})=0\nonumber\\
\frac{d \varphi}{d t}=0\nonumber
\end{eqnarray}
where $m^{c}_{i}$ is the constant part of magnetization components. From equation (\ref{syseq_simple}) it follows that $\varphi=const$, $I=\sin(\varphi-r m^{c}_{y})=I_{s}$, $m^{c}_{x}=0$, $m^{c}_{z}=\sqrt{1-(m^{c}_{y})^{2}}$ and, respectively,
\begin{eqnarray}
\label{my_const}
m^{c}_{y}&=&GrI_{s}
\end{eqnarray}

The obtained expression (\ref{syseq_simple}) explains the increase of $m_{y}^{max}$ with the growth of the average superconducting current, where the growth of $m_{y}^{max}$ is accompanied by oscillations of $m_{y}$ around $m_{y}^{c}$ due to the presence of Josephson oscillations. It also qualitatively agrees with the numerical results presented in Fig.\ref{Gdep}, which show the dependences of $m_{y}^{max}$ on $G$ and $r$.

\begin{figure}[h!]
\centering
\includegraphics[width=6cm]{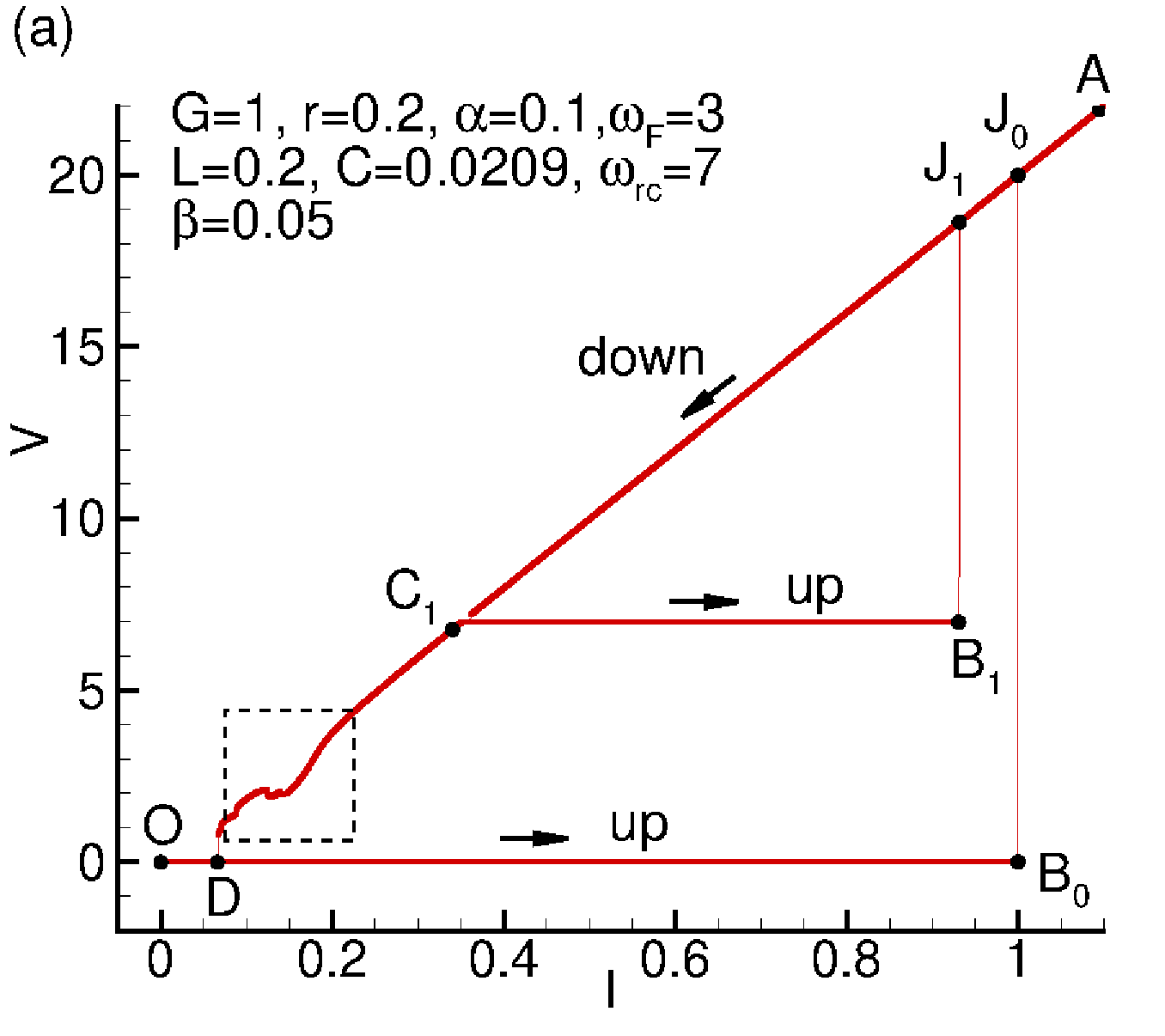}
\includegraphics[width=6cm]{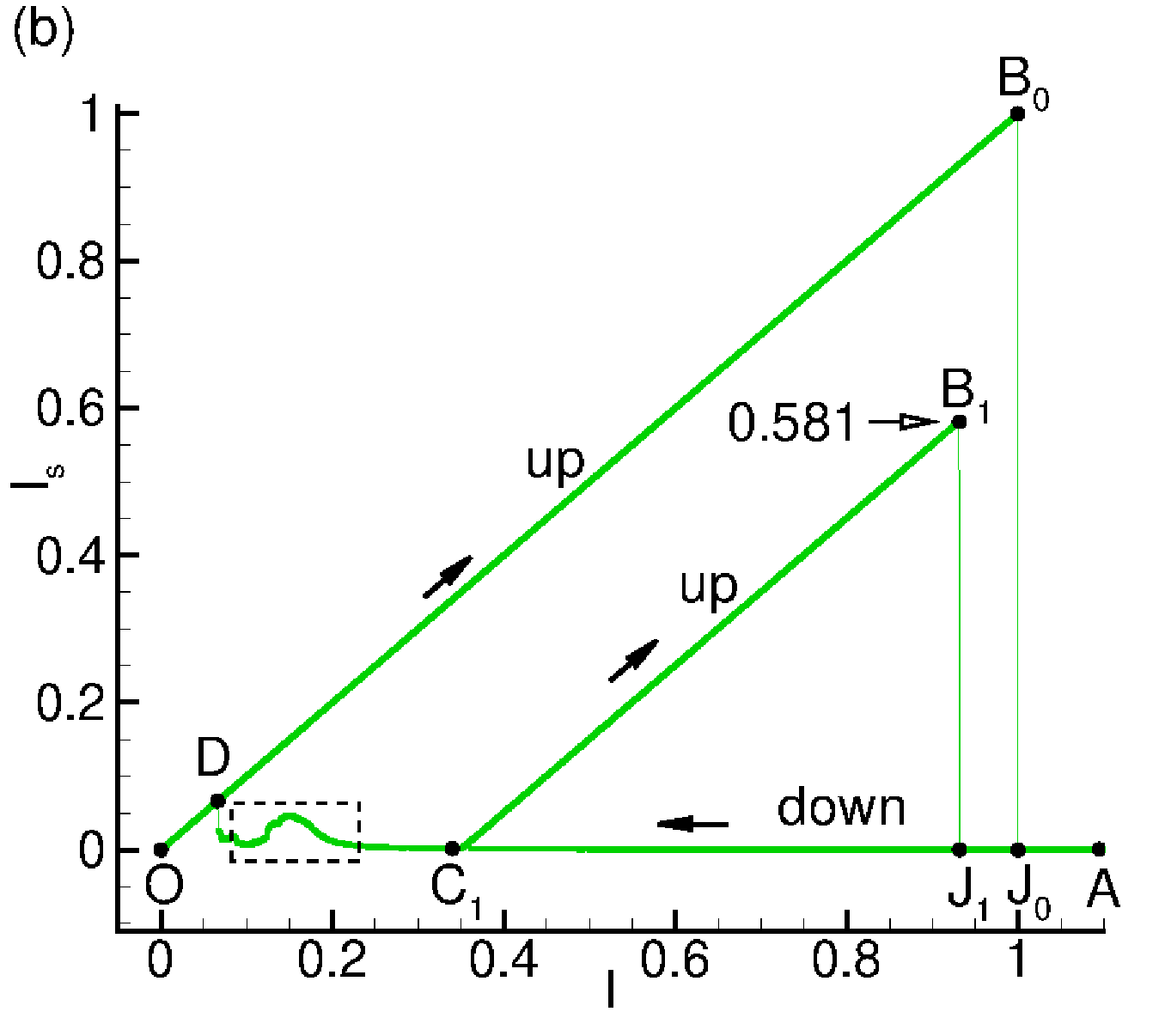}
\includegraphics[width=6cm]{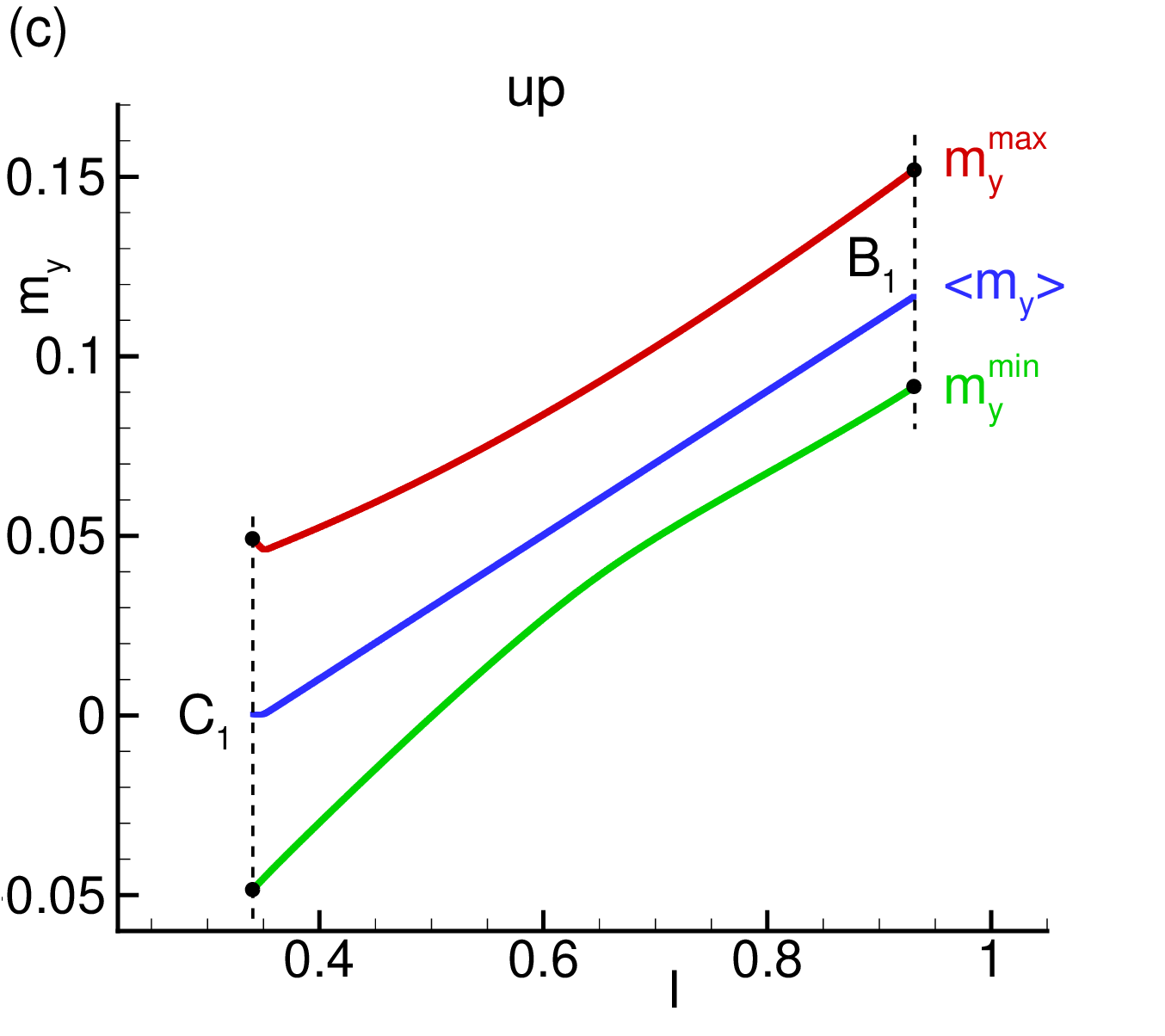}
\caption{(a) CVC of the $\varphi_{0}$ junction shunted by LC circuit, calculated at $\beta=0.05$, $\alpha = 0.1$, $G = 1$, $r = 0.2$, $\omega_{F}=3$, $C = 0.0209$, $L = 1$; (b) The dependence of average superconducting current on bias current obtained in the process of calculating CVC shown in Fig. (a); (c) Dependences of minimum, maximum, and average value of $m_{y}$ on bias current obtained  in the process of calculating CVC in the region of resonant branch with increasing current shown in Fig.\ref{cvc_ph0_shunt}}
\label{cvc_ph0_shunt}
\end{figure}

To check the quantitative agreement of the numerical results with expression (\ref{my_const}), we calculated the two-loop CVC and the dependence of the average value of the superconducting current of the shunted $\varphi_{0}$ junction on the bias current, which are shown in Fig.\ref{cvc_ph0_shunt}(a)-(b), respectively.

In the first loop (Fig.\ref{cvc_ph0_shunt}(a)) the current increases along the $OB_{0}A$ trajectory and decreases along $AC_{1}$. In the second loop, the current increases along $C_{1}B_{1}J_{1}$, and decreases along $J_{1}DO$; the $C_{1}B_{1}$ region corresponds to the resonant branch. We should like to note that the dip structure on CVC in the region marked with the dashed rectangle corresponds to the ferromagnetic resonance. Figure \ref{cvc_ph0_shunt}(b) demonstrates a linear increase in the superconducting current in the $OB_{0}$ region corresponding to the zero voltage state. In the second loop the calculation also shows an increase in the superconducting current in the region of the resonant branch $C_{1}B_{1}$. With a decrease in the bias current, a peak corresponding to ferromagnetic resonance is also observed here.

At the point $B_{1}$ corresponding to the maximum of the parallel resonance, the value of the average superconducting 
current is equal to $0.581$. Using this value and taking into account $G=1$, $r=0.2$, we can calculate $m_{y}^{c}$ by formula (\ref{my_const}) $m_{y}^{c}=1\cdot0.2\cdot0.581=0.1162$. Consequently, at the current value corresponding to the resonance peak the magnetization oscillates around $m_{y}^{c}=0.1162$. The dependence of the maximum, minimum, and average values of $m_{y}$ on the bias current during its increase along the CVC in the region of the resonant branch is shown in Fig.\ref{cvc_ph0_shunt}(c). With increasing the bias current in the first loop of the CVC in the section $OB_{0}$ corresponding to the stationary Josephson effect, the values $m_{y}^{max}$ and $m_{y}^{min}$ are equal to each other and increase linearly with the bias current, i.e. they agree well with expression (\ref{my_const}). The equality of $m_{y}^{max}$ and $m_{y}^{min}$ in this part of the CVC is due to the absence of Josephson oscillations.

In the second loop, with increasing bias current in the region $C_{1}B_{1}$, the values of $m_{y}^{max}$ and $m_{y}^{min}$ also increase but differ from each other in magnitude, which are demonstrate the oscillations of $m_{y}(t)$ around the average value $m_{y}^{c}$. The numerically calculated average value of $m_{y}$ at the resonance point is $0.1164$, while the analytical value obtained above is $0.1162$, i.e., there is a good agreement between the numerical and analytical results.

We emphasize that the oscillations of $m_{y}$ in the region of the resonance branch occur in the vicinity of its average value. Figure\ref{time_dep} shows the time dependences of $m_{x}$ (a), $m_{y}$ (b), and the magnetization trajectory on the $m_{x}-m_{y}$ plane (c). Analysis of the time dependences shows that $m_{x}(t)$ oscillates around the value $<m_x> = -0.005$ and $m_{y}(t)$ - around $<m_y> = 0.1164$. The dashed line in the time dependences shows the average value of the corresponding magnetization component. The magnetization trajectory on the $m_{x}-m_{y}$ plane demonstrates precession around the point $m_{x}=-0.005$, $m_{y}=0.1164)$, which confirms the fact of magnetization precession around the direction of the magnetic field created by the constant component of the superconducting current in the resonance region.

For visual demonstration of the magnetization deviation from the easy axis, in Fig.\ref{time_dep}(d) we show the positions of the magnetization vector on the $z$-$y$ plane for three values of the ratio of the Josephson energy to the magnetic one $G=1$, $G=2$ and $G=3$.

Additional confirmation of the reason for the increase in magnetization in the shunted $\varphi_{0}$ junction is presented in Fig.\ref{omega_dep}, which shows the dependences of the average values of the magnetization component $m^{av}_{y}$ and the superconducting current on the resonant circuit frequency. A strict correlation of these quantities is observed at different values of the resonance circuit frequency, which, in this case, was varied by changing the shunt inductance $L$ at a fixed value of the circuit capacitance ($C=0.02$).

Let us estimate the possibility of experimental realization of the observed effects. The values of the shunting capacitance can vary within wide limits, in particular, at $C_J=0.2 pF$ and inductance $L=50 pH$. The calculation of the dimensionless inductance leads to the value $L=2.5$. To observe the $rc$-branch on the CVC with these parameters and the equality $LC=0.5$, which determines the resonant frequency according to formula (\ref{w_par}), a shunting capacitance of $C_{sh}=0.04 pF$ is sufficient. Calculations of the CVC for JJ at different values of $L$ and $C$ are given in the works \cite{Shukrinov12,shukrinov_jetp17}. The observation of the $rc$-branch in the hysteresis region of the CVC at high inductance values makes it necessary to the need to use smaller values of the shunting capacitance. Some freedom in choosing the values of $L$ and $C$ can be provided by changing the area of the JJ, the plasma frequency, and also the choice of the ferromagnetic material.

An experimental investigation would require measuring $m_y(t)$ in a shunted $\varphi_0$ junction, which can be performed using DC-SQUID. A thin layer of $Pt$-doped permalloy \cite{Hrabec} on a dielectric substrate is proposed as a ferromagnet, where a sufficiently strong Rashba-type spin-orbit interaction is realized \cite{Shukrinov_apl2017}. Another suitable candidate could be a two-layer $Pt/Co$ ferromagnet without inversion symmetry, such as MnSi or FeGe. The spin-orbit interaction should generate a $\varphi_0$ Josephson junction with a finite fundamental phase difference. Measuring this phase difference (similar to the experiments in \cite{Szombati}) could serve as an independent way to estimate the parameter $r$. The parameter $G$ was estimated in the work \cite{konschelle-PRL}, and with a sufficiently strong anisotropy, one can expect $G\sim 1$.

Thus, the presented results demonstrate the possibility of controlling the magnetization in the shunted $\varphi_{0}$ junction by changing the resonant circuit parameters $L$ and $C$. Moreover, measuring $<m_y>$ and $I_s$ as functions the model parameters would allow the experimental determination of $r$ and $G$ according to equation (\ref{my_const}). We believe that the experimental implementation of such control will find application in modern superconducting electronics and spintronics.

In conclusion, we should like to note that we have demonstrated the possibility of resonant control of magnetization in a Josephson $\varphi_{0}$ superconductor-ferromagnet-superconductor junction shunted by an $LC$ circuit.

The resonance of Josephson oscillations with oscillations in the $LC$ circuit that occurs during shunting leads to the occurrence of a time-independent superconducting current and a deviation of the easy axis from its initial position. An interesting possibility arises of influencing the precession of magnetization by changing the parameters of the $\varphi_{0}$ junction and the $LC$ circuit. The obtained agreement between the numerical and analytical calculations creates a good basement for further theoretical and experimental investigations in this direction and can be used for the development of new technologies in the field of superconducting electronics and spintronics.

{\it Acknowledgment}: This work performed with the financial support of Russian Science Foundation in the framework of project 24-21-00340. We thank the Laboratory of Information Technologies of JINR for the opportunity to use the computational resources of the HybriLIT Heterogeneous Platform.

\end{document}


\title{Supplementary Material to ``Resonant control of magnetization in a shunted $\varphi_0$ junction''}

\author{I. R. Rahmonov}
\affiliation{BLTP, JINR, Dubna, Moscow Region, 141980, Russia}
\affiliation{Dubna State University, Dubna,  141980, Russia}

\author{Yu. M. Shukrinov}
\affiliation{BLTP, JINR, Dubna, Moscow Region, 141980, Russia}
\affiliation{Dubna State University, Dubna,  141980, Russia}

\author{O. Kibardina}
\affiliation{Dubna State University, Dubna,  141980, Russia}

\author{S. Abdelmoneim}
\affiliation{Menofia University, Cairo, Egypt}

\date{\today}
\maketitle

\begin{widetext}

\section{System of equations}
Coupled system of equations in normalized variables has a form

\begin{eqnarray}
\label{syseq}
\frac{d m_{x}}{d t}&=&-\frac{\omega_{F}}{1+\alpha^2m^2}\{(m_{y}h_{z}-m_{z}h_{y})-\alpha[m_{x}(m_{x}h_{x}+m_{y}h_{y}+m_{z}h_{z})-h_{x}m^2)]\}\nonumber
\vspace{0.5cm}\nonumber\\
\frac{d m_{y}}{d t}&=&-\frac{\omega_{F}}{1+\alpha^2m^2}\{(m_{z}h_{x}-m_{x}h_{z})-\alpha[m_{y}(m_{x}h_{x}+m_{y}h_{y}+m_{z}h_{z})-h_{y}m^2)]\}
\vspace{0.5cm}\nonumber\\
\frac{d m_{z}}{d t}&=&-\frac{\omega_{F}}{1+\alpha^2m^2}\{(m_{x}h_{y}-m_{y}h_{x})-\alpha[m_{z}(m_{x}h_{x}+m_{y}h_{y}+m_{z}h_{z})-h_{y}m^2)]\}
\vspace{0.5cm}\nonumber\\
\frac{d V}{d t}&=&I-\beta\bigg(\frac{d\varphi}{dt}-r\frac{d m_{y}}{d t}\bigg)-\sin(\varphi-r m_{y})-C\frac{du_{C}}{dt}\nonumber\\
\frac{d u_{c}}{dt}&=&U\nonumber\\
\frac{dU}{dt}&=&\frac{1}{LC}(V-u_{c})\nonumber\\
\frac{d \varphi}{d t}&=&V\nonumber
\end{eqnarray}
where $m_{i}$ are the magnetization components normalized to $M_{0}$, $t$ is the dimensionless time normalized to $\omega_{p}^{-1}$, $\omega_{p}=\sqrt{2eI_{c}/(\hbar C)}$ is the Josephson plasma frequency, $\omega_{F}=\Omega_{F}/\omega_{p}$ is the ferromagnetic resonance frequency, $\Omega_{F}=\gamma K/M_{0}$ normalized to $\omega_{p}$, $\beta$ is the dissipation parameter of the JJ, $C=C_{sh}/C_{J}$ is the normalized shunting capacitance, $L=L_{sh}(\omega_{p}^{2}C_{J})$ is the normalized shunting inductance, $I$ is the base current normalized to $I_{c}$, $V$ and $u_{c}$ are the voltages in the JJ and shunting capacitance normalized to $V_{0}=\hbar\omega_{p}/2 e$, respectively. Here $h_{x}$, $h_{y}$ and $h_{z}$ denote the components of the effective field normalized to $M_{0}/K$ and are defined by the expressions

\begin{eqnarray}
\label{eff_field}
h_{x}&=&0,\nonumber
\vspace{0.5cm}\\
h_{y}&=&rG\sin(\varphi-rm_{y}),
\vspace{0.5cm}\\
h_{z}&=&m_{z}.\nonumber
\end{eqnarray}

\section{Effective field}

The expression for effective field can be found as  

\begin{eqnarray}
\label{heff1}
\mathbf{H_{eff}}=-\frac{1}{\nu}\frac{\delta E_{t}}{\delta\mathbf{M}},
\end{eqnarray}
where $E_{t}$  is total energy of system, $\nu$ volume of ferromagnet layer.

In our case, the total energy of the system consists of the energy of the external current $-\frac{\Phi_{0}}{2\pi}I\varphi$, the potential energy of the Josephson junction $E_{s}(\varphi,\varphi_{0})$ and the energy of magnetic anisotropy
 $E_{M}(M)$
\begin{eqnarray}
\label{Et}
E_{t}=-\frac{\Phi_{0}}{2\pi}I\varphi+E_{s}(\varphi,\varphi_{0})+E_{M}(\mathbf{M}),
\end{eqnarray}
where $\Phi_{0}=h/2e$ is the magnetic flux quantum, $I$ is the base current. The potential energy of the Josephson junction is determined by the expression
\begin{eqnarray}
\label{Es}
E_{s}=(\varphi,\varphi_{0})=E_{J}[1-\cos(\varphi-\varphi_{0})],
\end{eqnarray}
where $\varphi_{0}=rM_{y}/M_{0}$, $r=l\upsilon_{SO}/\upsilon_{F}$ is the spin-orbit interaction parameter. The expression for the magnetic anisotropy energy is

\begin{eqnarray}
\label{Em}
E_{M}(\mathbf{M})=-\frac{K\nu}{2}\bigg(\frac{M_{z}}{M_{0}}\bigg)^{2},
\end{eqnarray}
where $K$ is the anisotropy constant.

\begin{eqnarray}
\label{heff2}
\mathbf{H_{eff}}=\frac{K}{M_{0}}\bigg[Gr\sin\bigg(\varphi-r\frac{M_{y}}{M_{0}}\bigg)\mathbf{e_{y}}+\frac{M_{z}}{M_{0}}\mathbf{e_{z}}\bigg]
\end{eqnarray}
where $\displaystyle G=\frac{E_{J}}{K\nu}$ is the ratio of the amplitude of the Josephson energy to the magnetic one, $\mathbf{e_{y}}$ and $\mathbf{e_{z}}$ are unit vectors.

\section{Multi loop CVC of shunted SIS junction}

Here we have described the calculation method of multi loop CVC in example of shunted SIS junction, which presentd in Fig.\ref{shunted_sis_jj}(a).
\begin{figure}[h!]
\centering
\includegraphics[width=8.5cm]{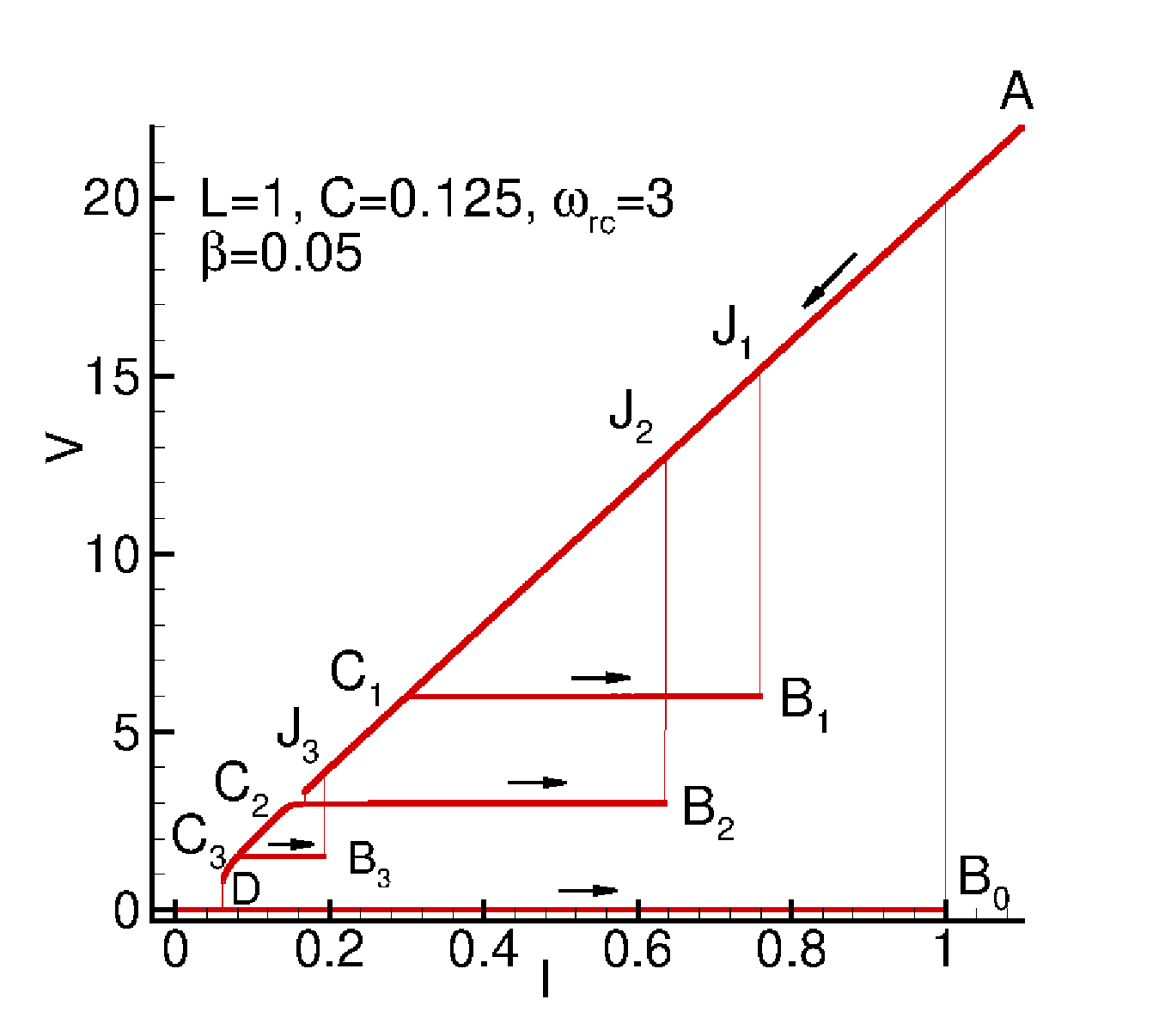}
\caption{ The I-V characteristic of the SIS junction shunted by an LC circuit, calculated at $\beta=0.05$, $C=0.125$, $L=1$. The arrows indicate the direction of change of the base current, and the symbols indicate the current values corresponding to the beginning, end and points of change of the I-V characteristic}
\label{shunted_sis_jj}
\end{figure}

The calculation was performed with the dissipation parameter $\beta=0.05$, shunt capacitance $C=0.125$, inductance $L=1$, which lead to the natural frequency $\omega_{rc}=3$. In the figure, the arrows indicate the direction of bias current changing, and the symbols indicate the points corresponding to the beginning, end and points of change of the I-V characteristic. The calculation starts from point $O$ along the branch with zero voltage to the critical current $I=1$, i.e. to point $B_{0}$. Then there is a jump to the resistive branch and an increase in current to $I_{max}$ (point A) with its subsequent decrease. The I-V characteristic shown in Fig.\ref{shunted_sis_jj}(a) is the result of a fourfold increase and decrease of the base current: the first loop corresponds to the trajectory $OB_{0}AC_{1}$, the second - $C_{1}B_{1}J_{1}C_{2}$, the third - $C_{2}B_{2}J_{2}C_{3}$ and finally the fourth - $C_{3}B_{3}J_{3}DO$. Thus, the branch corresponding to the main resonance is designated $C_{2}B_{2}$, the second harmonic - $C_{1}B_{1}$ and the subharmonic $C_{3}B_{3}$.

The reason for the formation of these branches is the occurrence of a time-independent superconducting current through the JJ as a result of the resonance of Josephson oscillations and with oscillations in the LC circuit.

\end{widetext}